\documentclass[9pt,twocolumn,twoside]{pnas-new}
\articletype{}

\templatetype{pnasresearcharticle} 

\usepackage{bm}

\let\OriginalJournalTitle\JournalTitle
\renewcommand{\JournalTitle}[1]{\ifstrempty{#1}{}{\OriginalJournalTitle{#1}}}
\patchcmd{\endthebibliography}{\label{LastPage}}{}{}{}
\setboolean{displaywatermark}{false}
\begin{document}

\title{Designing tunable hyperuniformity in random organization through nonlocal noise}
\author[a]{Yusheng Lei}
\author[a]{Yuhan Peng}
\author[a,1]{Ran Ni}
\affil[a]{School of Chemistry, Chemical Engineering and Biotechnology, Nanyang Technological University, 62 Nanyang Drive, Singapore 637459}
\leadauthor{Lei}
\significancestatement{Disordered hyperuniform systems suppress large-scale density fluctuations without forming a crystal. We show how microscopic particle dynamics can select the scaling of this suppression while preserving ordinary diffusion and pairwise center-of-mass conservation. Two mechanisms connect collective structure to the spatial statistics of driving: density sensing of a surrounding Brownian gas and reciprocal particle kicks with broadly distributed lengths. In the latter, the tail of the kick distribution directly selects the hyperuniform exponent, although particle motion remains diffusive. These results provide a microscopic design principle for controlling large-scale correlations in disordered particle systems.
}
\authorcontributions{Author contributions: R.N. and Y.L. designed research; Y.L. performed research; Y.L., Y.P. and R.N. analyzed data; and Y.L., Y.P. and R.N. wrote the paper.}
\authordeclaration{Competing interests: The authors declare that they have no competing interests.}
\correspondingauthor{\textsuperscript{1}To whom correspondence should be addressed. E-mail: r.ni@ntu.edu.sg}

\begin{abstract}
Hyperuniform fluids suppress density fluctuations at large length scales, but microscopic protocols for designing their asymptotic scaling remain limited. Here we show how effective driving noise generated by particle dynamics can select the hyperuniform exponent while preserving ordinary diffusion and pairwise center-of-mass conservation of the organized particles. We establish this connection through two protocols based on biased random organization (BRO). First, a two-dimensional BRO layer is nonreciprocally coupled to a three-dimensional Brownian gas through density-dependent kick amplitudes. Bulk density fluctuations generate spatially nonlocal noise with a $q^3$ power spectrum in the adiabatic limit, producing $S(q\to0)\sim q^1$. Second, we introduce reciprocal kicks whose lengths follow a Pareto-like distribution, $P(\ell)\sim\ell^{-1-b}$. For $2<b<4$, the finite second moment supports ordinary diffusion, while the divergent fourth moment changes the long-wavelength noise spectrum to $q^b$. The corresponding fluctuating density field theory predicts continuously tunable hyperuniform scaling, $S(q\to0)\sim q^{b-2}$, supported by particle simulations. Together, these results connect targeted hyperuniform scaling to explicit microscopic dynamics and provide a physical basis for designing collective density correlations in nonequilibrium particle systems.
\end{abstract}
\dates{This manuscript was compiled on \today}

\maketitle
\thispagestyle{firststyle}
\ifthenelse{\boolean{shortarticle}}{\ifthenelse{\boolean{singlecolumn}}{\abscontentformatted}{\abscontent}}{}
\Firstpage

Disordered hyperuniform structures exhibit anomalously suppressed density fluctuations at large length scales, characterized by the structure factor $S(q\to0)=0$~\cite{Torquato2003,Torquato2018}, and they have been found in various systems, including avian photoreceptor patterns, jammed sphere packings, looped leaf vein networks, sheared colloidal suspensions, random-organizing systems, cellular structures, and chiral active fluids~\cite{Jiao2014,Donev2005,Hexner2015,Lei2019Chiral,Wilken2020,lei2023random,lei2023howHUfreeze,kuroda2023microscopic,Liu2024,zheng2024universal,li2025fluidization,lei2026metastable,torquato2026hyperuniformity}.
Such structures provide a basis for designing disordered photonic materials with band gaps and transparent disordered media~\cite{Florescu2009,Leseur2016}, and these findings motivate the controlled design of spatial correlations in disordered materials. Specifically, for disordered hyperuniform systems with $S(q\to0)\sim q^\alpha$, the exponent $\alpha>0$ specifies how rapidly the structure factor vanishes at long wavelengths and provides a natural target for controlling large-scale density fluctuations.

Various numerical methods have been devised to generate static hyperuniform structures with prescribed spectral densities~\cite{uche2006collective,ChenTorquato2018,shi2023computational,shi2025three}. For interacting multi-particle systems, targeted pair correlations can also be constructed in equilibrium by designing effective pair interactions~\cite{Zhang2020Realizable,Wang2024Designer}. At nonzero temperature, equilibrium hyperuniform states generally require long-range interactions, as in Coulomb plasmas~\cite{Lomba2017}, with their scaling determined by the long-range form of the interaction potential~\cite{Wang2024Designer}. However, experimental realization of the designed long-range interactions remains highly challenging. 

\Endparasplit

Out of equilibrium, hyperuniform structures can emerge through particle dynamics with short-range interactions~\cite{Lei2019Hydro,lei2025non}. Theoretical studies have explored how conservation laws and noise statistics influence hyperuniformity~\cite{Hexner2017,Maire2025,Ikeda2024}. A basic starting point for designing the hyperuniform scaling is the relation between diffusion, driving noise, and density fluctuations. Diffusion smooths density inhomogeneities, while fluctuations in particle motion and driving source continually generate them. For a density fluctuation field in spatial Fourier $\boldsymbol{q}$-space obeying $\partial_t\delta\tilde\rho(\boldsymbol q,t)=-\tilde F(\boldsymbol q)\delta\tilde\rho(\boldsymbol q,t)+\tilde\xi(\boldsymbol q,t)$, with temporally white driving noise, the steady-state structure factor is~\cite{Hexner2017}
\begin{equation}
 S(q)=\frac{\tilde C(q)}{2\bar\rho\,\tilde F(q)}.
 \label{eq:structure}
\end{equation}
Here, $\bar\rho$ is the mean particle density, $q=|\boldsymbol q|$ is the wavenumber, $\tilde F(q)$ describes the deterministic diffusion, and the noise spectrum $\tilde C(q)$ is defined by $\langle\tilde\xi(\boldsymbol q,t)\tilde\xi^*(\boldsymbol q',t')\rangle=(2\pi)^2\tilde C(q)\delta(\boldsymbol q-\boldsymbol q')\delta(t-t')$. This relation provides a basic principle: designing spatially nonlocal driving noise allows the hyperuniform scaling to be tuned while preserving ordinary diffusion. However, the challenge remains to turn this principle into physical protocols that generate the required noise correlations through particle motion and interactions. 
Particle models in nonequilibrium hyperuniform steady states typically exhibit quadratic scaling $S(q\to0)\sim q^2$~\cite{Lei2019Hydro,Hexner2017}.
Although nonquadratic hyperuniform scaling has been reported in systems with long-range hydrodynamic interactions~\cite{Huang2021,Oppenheimer2022}, a systematic microscopic route to continuously tuning the asymptotic exponent in disordered nonequilibrium steady states remains an open challenge. 


\begin{figure*}[t]
 \centering
 \includegraphics[width=\linewidth]{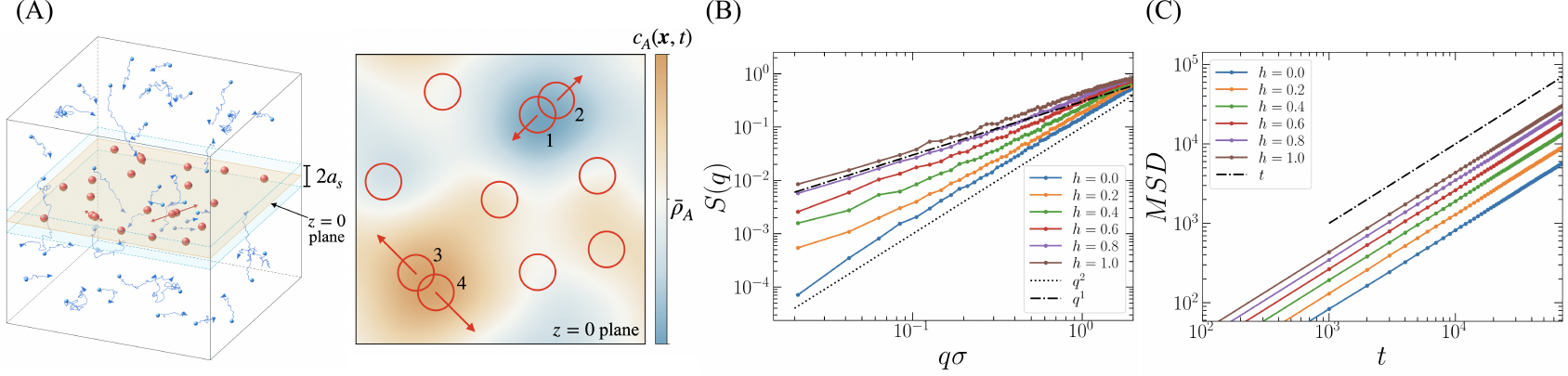}
 \caption{{\bf BRO with nonreciprocal coupling across dimensions.} (A) Schematic of a three-dimensional Brownian ideal gas (blue particles with trajectory lines) nonreciprocally coupled to BRO particles (red particles) confined in the $z=0$ plane (orange plane). The blue region depicts the approximated sensing window around it. The schematic heat map shows the sensed A density $c_A(\boldsymbol x,t)$: blue and orange indicate densities below and above the mean $\bar\rho_A$, respectively. Red circles represent BRO particles, and arrows indicate the equal-and-opposite kicks of overlapping pairs. Pair $(3,4)$ experiences enhanced kicks in the high-sensed density region, whereas pair $(1,2)$ receives the intrinsic kick strength set by $h_0$. (B) Structure factor $S(q)$ for systems of B particles with various $h$. As $h$ increases from $0$ to $1$, the $q^1$ scaling regime broadens and extends to larger $q$. (C) The mean-squared displacement (MSD) for the systems in (B) remains linear at long time, with increasing diffusivity due to the larger average kick amplitude. The simulation box we used here is a cubic box of size $L$ with periodic boundary conditions in all directions. Parameters are $L=300\sigma$, $a_s=2\sigma$, $\bar\rho_A=0.6\sigma^{-3}$, $\bar\rho_B=0.6\sigma^{-2}$, $h_0=1$, and $D_A=1$, with $h$ the coupling strength.}
 \label{fig:bulk}
\end{figure*}

In this work, we propose two physical protocols based on the center-of-mass-conserving biased random
organization (BRO) model. Our first protocol nonreciprocally couples a two-dimensional center-of-mass-conserving BRO layer to a three-dimensional Brownian gas through density sensing. The local density fluctuations enhance the particle kick amplitudes and generate spatially correlated driving noise in the plane, leading to $S(q \to 0)\sim q^1$. The second protocol introduces heavy-tailed reciprocal kicks directly into the BRO model. Changing the statistics of long jumps continuously varies the hyperuniform scaling $S(q \to 0)\sim q^\alpha$ over $0<\alpha<2$. Through particle simulations and fluctuating density field theories, we establish how these microscopic dynamics generate spatially nonlocal noise and control hyperuniform scaling while preserving ordinary diffusion and center-of-mass conservation. These protocols connect the  design principle to explicit particle dynamics and provide a physical starting point for exploring realizations of hyperuniform states with tunable scaling in nonequilibrium systems.

\section*{BRO with nonreciprocal coupling across dimensions}
\label{sec:bulk}
The original random organization model was introduced to describe the reversible--irreversible transition in periodically driven suspensions~\cite{pine2005chaos,Corte2008}. At each step, particles are checked for overlap. Overlapping particles are regarded as active particles and receive random displacements, while particles without overlap remain immobile. Here we use a center-of-mass-conserving BRO model, in which overlapping pairs receive equal-and-opposite displacements along their line of centers, directed away from one another~\cite{Hexner2017}. 
In this work, we introduce a nonreciprocal density-sensing coupling between a three-dimensional Brownian ideal gas and a two-dimensional BRO layer, as shown in Fig.~\ref{fig:bulk}(A). Species A is a Brownian ideal gas moving in the bulk, whose local density is sensed by species B, the BRO particles of diameter $\sigma$ confined to an immersed plane. The kick length of an overlapping B pair is drawn uniformly from $[0, h_0+h\Theta ( c_A(\boldsymbol R,t)-\bar\rho_A ) ]$, where $\boldsymbol R$ is the pair midpoint, $\Theta$ is the Heaviside step function, $h$ is the microscopic coupling strength, and $h_0$ is the intrinsic kick magnitude parameter. Here $c_A$ is the sensed local density of A near the midpoint $\boldsymbol R$ of an overlapping B pair, measured using a spherical window of small radius $a_{\mathrm s}=2\sigma$ in the simulations. Both B particles in an overlapping pair use the same amplitude, so each pair displacement conserves their center of mass. We neglect collisions between A and B, and the coupling is nonreciprocal: the sensed density of A changes the kick amplitudes of B, while B has no impact on A. We choose $h_0$ large enough to keep B active even without A. In the simulations, we take the interval between successive BRO updates as the unit of time, $\tau=1$. The Brownian ideal gas uses the same time step, $dt=\tau$, and both species are updated synchronously.

In Fig.~\ref{fig:bulk}(B), we plot the structure factor of B for different coupling strengths $h$. We find that, compared with the uncoupled system at $h=0$, the coupled system at $h=1$ shows a weaker hyperuniform scaling, i.e., $S(q \to 0)\sim q^1$. As $h$ increases, the $q^1$ scaling regime broadens and extends to larger $q$. The mean-squared displacement (MSD) remains proportional to time, with a diffusivity that increases with $h$ (Fig.~\ref{fig:bulk}(C)), originating from the increase in the average kick amplitude.

To understand this change in hyperuniform scaling, we construct a fluctuating density field theory for the coupled system~\cite{Dean1996,Hexner2017,Peng2026}.
We consider homogeneous active states away from the absorbing transition, with both the bulk and the plane infinitely large in the thermodynamic limit. The density fields of A and B are $\rho_A(\boldsymbol x,z,t)$ and $\rho_B(\boldsymbol x,t)$, respectively, where $\boldsymbol x=(x,y)$. Approximating the sensing region by a slab of thickness $2a_{\mathrm s}$ centered at plane $z=0$, we write
\begin{equation}
\begin{aligned}
 \partial_t\rho_A(\boldsymbol x,z,t)
 &=D_A\nabla_3^2\rho_A+
 \nabla_3\cdot\left[\sqrt{2D_A\rho_A}\boldsymbol\eta_1\right],\\
 c_A(\boldsymbol x,t)&=\int_{-\infty}^{\infty}dz\,
 W(z,a_{\mathrm s})\rho_A(\boldsymbol x,z,t),\\
 \partial_t\rho_B(\boldsymbol x,t)
 &=\nabla^2[D_B(c_A)\rho_B]
 +\nabla^2\left[\sqrt{\rho_B}B(c_A)\eta_2\right].
\end{aligned}
\label{eq:bulk-full-equation}
\end{equation}
Here $D_A$ is the bulk diffusion coefficient. $\nabla_3\equiv(\nabla,\partial_z)$ and $\nabla\equiv(\partial_x,\partial_y)$ are the gradient operators in three dimensions and in the 2D $xy$-plane, respectively. $\boldsymbol\eta_1$ and $\eta_2$ are independent unit Gaussian white noises. The normalized sensing window is $W(z,a_{\mathrm s})=\Theta(a_{\mathrm s}-|z|)/(2a_{\mathrm s})$, giving the mean sensed bulk density $\bar c_A=\bar\rho_A$.
The first equation in Eq.~\ref{eq:bulk-full-equation} describes the dynamics of species A, an ideal Brownian gas. For species B, the second and third equations in Eq.~\ref{eq:bulk-full-equation} describe the density-sensing coupling, through which the local density of A changes the kick amplitude of each overlapping B pair without breaking the center-of-mass conservation. Within a local isotropic approximation, we retain this conservation through the Laplacian form of both terms, with the density-dependent coefficients inside the operators, which recovers ordinary diffusion when $D_B$ is constant.

We use the phenomenological linear coupling $D_B(c_A)=\bar D_B+h c_A$, where $\bar D_B$ is the intrinsic diffusivity and $h$ is a coarse-grained coupling coefficient that plays the same qualitative role as the microscopic coupling parameter in the simulations. Both describe how strongly the sensed density of A affects the motion of B. We then introduce the fluctuation fields $\delta\rho_A=\rho_A-\bar\rho_A$, $\delta\rho_B=\rho_B-\bar\rho_B$, and $\delta c_A=c_A-\bar c_A$. Linearizing the deterministic terms and evaluating the noise amplitudes at the mean densities, we have (see \textit{SI Appendix})
\begin{equation}
\begin{aligned}
\partial_t\delta\rho_A&=D_A\nabla_3^2\delta\rho_A +\sqrt{2D_A\bar\rho_A}\nabla_3\cdot\boldsymbol\eta_1,\\
 \partial_t\delta\rho_B&=\bar D_{B,\mathrm{eff}}\nabla^2\delta\rho_B
 +h\bar\rho_B\nabla^2\delta c_A
 +\bar B_{\mathrm{eff}}\sqrt{\bar\rho_B}\nabla^2\eta_2,
\end{aligned}
\label{eq:coupled}
\end{equation}
where $\bar D_{B,\mathrm{eff}}=\bar D_B+h\bar\rho_A$ and $\bar B_{\mathrm{eff}}=B(\bar\rho_A)$. The first two terms on the right hand side of the species $B$ equation come from the activity-driven diffusion $\nabla^2\left[D_B(c_A )\rho_B\right]$, which preserves the center of mass of the system of B particles.

By solving Eq.~\ref{eq:coupled} in Fourier space, we obtain the structure factor (see \textit{SI Appendix}):
\begin{equation}
 S_B(q\to 0)\simeq
 \frac{h^2\bar\rho_A\bar\rho_Bq}
 {2D_A\bar D_{B,\mathrm{eff}}\sqrt{1+\bar D_{B,\mathrm{eff}}/D_A}}+O(q^2),
 \label{eq:slabSk}
\end{equation}
where first term comes from the coupling to A, while the second term is the usual contribution from the Laplacian noise of B themselves and the correction due to the finite sensing width (see \textit{SI Appendix}). For any nonzero coupling, the $q^1$ term dominates at $q\to 0$. Varying $h$ changes its amplitude and the crossover scale, rather than continuously tuning the asymptotic exponent. This model nevertheless shows that center-of-mass conservation permits $q^1$ scaling through nonreciprocal coupling across dimensions.

According to Eq.~\ref{eq:coupled}, the effect of A on B can be regarded as an effective noise in the form of $\xi=\nabla^2\zeta$, where $\zeta=h\bar\rho_B\delta c_A$. Since A particles diffuse through the sensing region, this noise is spatiotemporally correlated (see \textit{SI Appendix}). Because of its diffusive dynamics, B responds mainly to fluctuations of A at frequencies $|\omega|\lesssim\bar D_{B,\mathrm{eff}}q^2$. In the adiabatic limit $D_A\gg\bar D_{B,\mathrm{eff}}$, we can therefore describe the effective noise as temporally white, with its strength determined by the zero-frequency spectrum of A. At large length scales $qa_{\mathrm s}\ll1$, the density fluctuation field equation then becomes
\begin{equation}
 \begin{aligned}
 \partial_t\delta\rho_B&\simeq\bar D_{B,\mathrm{eff}}\nabla^2\delta\rho_B+\xi
 +\bar B_{\mathrm{eff}}\sqrt{\bar\rho_B}\nabla^2\eta_2,\\
 \xi&=\nabla^2\zeta,\,\,\text{with}\,\,
 \zeta\simeq h\bar\rho_B\sqrt{\frac{\bar\rho_A}{D_A}}
 (-\nabla^2)^{-1/4}\eta.
 \end{aligned}
 \label{eq:adiabatic}
\end{equation}
where $\eta$ is Gaussian white noise. The effective noise is therefore a temporally white, spatially nonlocal fractional-Laplacian noise $\xi\simeq -h\bar\rho_B\sqrt{\bar\rho_A/D_A}\allowbreak(-\nabla^2)^{3/4}\eta$, whose spectrum is proportional to $q^3$, leading to $S_B(q \to 0)\simeq h^2\bar\rho_A\bar\rho_Bq/(2D_A\bar D_{B,\mathrm{eff}})$. Here, the Fourier definition of the fractional-Laplacian operator is employed \cite{kwasnicki2017ten,lischke2020fractional}.

\section*{Effective two-dimensional coupling description}
\label{sec:plane}
To understand how the bulk particles generate this nonlocal noise, we consider the limit $a_{\mathrm s}\to0$, and define the effective sensed planar density fluctuation field as $\delta c(\boldsymbol x,t)\equiv l_0 \lim_{a_{\mathrm s}\to0}\delta c_A(\boldsymbol x,t)=l_0\delta\rho_A(\boldsymbol x,z=0,t)$, where $l_0$ is the unit of length. Hereafter, quantities are expressed in the corresponding reduced units with $l_0=1$, and explicit factors of $l_0$ are omitted. The coupling in Eq.~\ref{eq:coupled} can then be described entirely in the plane, where the effective two-dimensional density fluctuation field $\delta c$ drives the density fluctuations of the BRO particles. Following an approach similar to that used for bulk-mediated surface diffusion~\cite{Chechkin2012}, we integrate out the motion perpendicular to the plane and obtain (see \textit{SI Appendix}):
\begin{equation}
 \sqrt{i\omega+D_Aq^2}\,\delta\tilde c
 =\sqrt{\bar\rho_A}
 \left[\frac{\sqrt{q^4+\omega^2/D_A^2}+q^2}{2}\right]^{1/4}\tilde\eta.
 \label{eq:plane}
\end{equation}
where, in real space, the deterministic part of the effective equation involves the fractional heat operator $(\partial_t-D_A\nabla^2)^{1/2}$ in 2D, and is nonlocal in both time and space~\cite{stinga2017regularity}, because the density in the plane retains the memory of diffusion in the bulk. A similar square-root form also appears in the  bulk-mediated surface diffusion systems~\cite{Bychuk1995,Chechkin2012}. In the low-frequency limit $|\omega|\ll D_Aq^2$, we find $\delta\tilde c(q,\omega)\simeq[\bar\rho_A/(D_Aq)]^{1/2}\tilde\eta(q,\omega)$. Substituting this into $\xi=h\bar\rho_B\nabla^2\delta c$ recovers the effective nonlocal noise in Eq.~\ref{eq:adiabatic}. The effective planar field thus shares the $q^{-1}$ scaling of the zero-frequency density fluctuation spectrum of a 2D ideal gas undergoing Cauchy flights, a special case of L\'evy flights with stability index $\mu=1$ \cite{Metzler2007}. Its full dynamics retains the diffusive time scale of the bulk, with a density correlation proportional to $t^{-1/2}\exp(-D_Aq^2t)$ for $t>0$. The power-law term reflects the memory of bulk excursions, which can appear as long jumps between distant locations on the plane, while the exponential term describes effective long-time ordinary diffusion along the plane. The bulk therefore generates Cauchy-like low-frequency fluctuations while retaining diffusive density correlations. A related distinction also appears in bulk-mediated surface diffusion~\cite{Chechkin2012}, where the surface propagator crosses over from a Cauchy-type form to a Gaussian form at long times, with the normalized surface mean-squared displacement growing linearly with time.

\begin{figure*}[t]
 \centering
 \includegraphics[width=\linewidth]{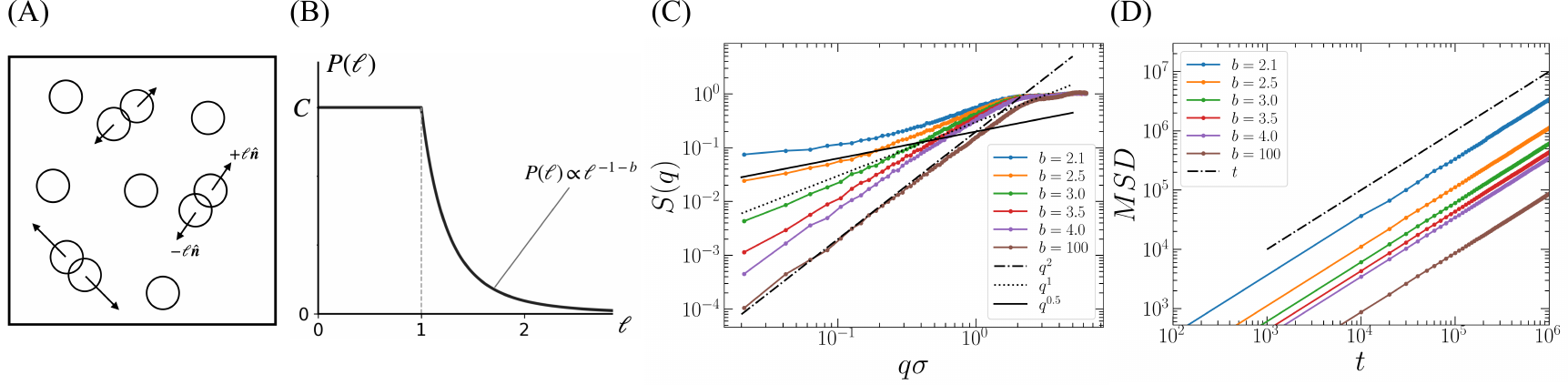}
 \caption{{\bf BRO models with heavy-tailed kicks.} (A) Overlapping particles receive reciprocal displacements along the line connecting their centers and away from its midpoint, denoted as $+\ell\hat{\boldsymbol{n}}$ and $-\ell\hat{\boldsymbol{n}}$ with unit vector $\hat{\boldsymbol{n}}$. (B) The probability density distribution of kick-length $P(\ell)$, which has a uniform core and a Pareto-like tail. (C) Structure factors for $b=2.1,2.5,3,3.5,4,100$, with reference scaling $q^{0.5}$, $q^1$, and $q^2$. The theory predicts $S(q \to 0)\sim q^{b-2}$ for $2<b<4$ and $q^2$ for $b>4$, while $S(q \to 0)\sim q^2\ln[1/(q\sigma)]$ for $b=4$. (D) MSD for the systems in (C), which remains linear at long time. The simulation box we used here is a square box of size $L$ with periodic boundary conditions in all directions. Parameters are $L=300\sigma$ and $\bar\rho=0.6\sigma^{-2}$.}
 \label{fig:kicks}
\end{figure*}

Although coupling to Brownian particles in 3D fixes the hyperuniform exponent to one, the effective planar description suggests possibilities to tune the effective noise spectrum by changing the dynamics of species A. As a further generalization, suppose that A is a two-dimensional ideal gas performing symmetric L\'evy flights, with mean density $\bar\rho_A$ (see \textit{SI Appendix}). In Fourier space, its density fluctuation field obeys
\begin{equation}
    \partial_t\delta\tilde\rho_A(\boldsymbol q,t)=-D_Aq^\mu\delta\tilde\rho_A+\sqrt{2D_A\bar\rho_Aq^\mu}\tilde\eta_A,
\end{equation}
where the first term describes fractional diffusion with generalized diffusion coefficient $D_A$, $\mu$ is the stability index~\cite{Fogedby1994,Jespersen1999}, and $\eta_A$ is a unit Gaussian white noise. For $0<\mu<2$, the L\'evy jump-length distribution has a tail $\Delta x^{-1-\mu}$, so larger $\mu$ corresponds to a lighter tail and weaker superdiffusion. Ordinary diffusion is recovered at $\mu=2$. At $|\omega|\ll D_Aq^\mu$, its density fluctuation field is $\delta\tilde\rho_A(q,\omega)\simeq[2\bar\rho_A/(D_Aq^\mu)]^{1/2}\tilde\eta_A(q,\omega)$. With the same density-sensing coupling but both species in 2D, we obtain in the adiabatic limit (see \textit{SI Appendix}):
\begin{equation}
    S_B(q \to 0)\simeq\frac{h^2\bar\rho_B\bar\rho_A}{\bar D_{B,\mathrm{eff}}D_A}
  q^{2-\mu}+\frac{\bar B_{\mathrm{eff}}^2}{2\bar D_{B,\mathrm{eff}}}q^2.
\end{equation} 
Thus $S_B(q\to 0)\sim q^{2-\mu}$ for $0<\mu<2$, while ordinary diffusion at $\mu=2$ generates a nonhyperuniform plateau. The role of long jumps in this construction motivates introducing them directly into BRO in our next approach.

\section*{BRO with heavy-tailed kicks}
\label{sec:heavy}
Next, we introduce long jumps directly into the center-of-mass conserving BRO dynamics in two dimensions. At each simulation step, every overlapping pair receives reciprocal displacements $\pm\ell\hat{\boldsymbol n}$ along its line of centers and away from its midpoint, and a particle may participate in several pair kicks. The kick length $\ell$ is drawn from a Pareto-like distribution,
\begin{equation}
 P(\ell)=\begin{cases}
 C,&0\leq\ell\leq\sigma,\\
 C(\ell/\sigma)^{-1-b},&\ell>\sigma,
 \end{cases}
 \qquad b>2,
 \label{eq:pareto}
\end{equation}
where $C=b/[\sigma(1+b)]$ normalizes the distribution. This modification preserves the center of mass for every pair displacement, while allowing occasional large kicks (Fig.~\ref{fig:kicks}(A,B)). As shown in Fig.~\ref{fig:kicks}(C), the exponent in hyperuniform scaling $S(q\to 0)\sim q^{b-2}$ increases with increasing $b$ for $2<b<4$. Despite these long jumps, the particle motion remains diffusive for $b>2$, as indicated by the MSD in Fig.~\ref{fig:kicks}(D). In the limit $b\to\infty$, the tail disappears and the distribution becomes uniform on $[0,\sigma]$, recovering the conventional center-of-mass-conserving BRO model and $S(q \to 0) \sim q^2$ scaling. Our results indicate that long kicks move particles between distant regions, weakening the local cancellation produced by reciprocal displacements without making the overall motion superdiffusive for $b>2$.

To understand how the kick distribution changes the dynamics and structure factor, we write the density fluctuation field equation in Fourier space as $\partial_t\delta\tilde\rho(\boldsymbol q,t) =-\tilde F(q)\delta\tilde\rho(\boldsymbol q,t) +\tilde\xi(\boldsymbol q,t)$, where $\boldsymbol q=(q_x,q_y)$ and $\tilde F(q)$ determines the diffusion term. The driving noise has covariance $\langle\tilde\xi(\boldsymbol q,t)\tilde\xi^*(\boldsymbol q',t')\rangle=(2\pi)^2\tilde C(q)\delta(\boldsymbol q-\boldsymbol q')\delta(t-t')$, where $\tilde C(q)$ is the noise spectrum. We determine $\tilde F(q)$ and $\tilde C(q)$ from the mean and variance of the density increment generated by pair kicks over a short time interval $\Delta t$. For simplicity, we neglect the initial overlapping pair separation and represent both particles at their midpoint. Thus the density change caused by a pair-kick event $e$ can be approximated by $g_e(\boldsymbol x)=\delta(\boldsymbol x-\boldsymbol\ell)+\delta(\boldsymbol x+\boldsymbol\ell)-2\delta(\boldsymbol x)$, with $\boldsymbol\ell=\ell\hat{\boldsymbol n}$, and the density increment is $\Delta\rho(\boldsymbol x,t)\equiv\rho(\boldsymbol x,t+\Delta t)-\rho(\boldsymbol x,t)=\sum_{e\in[t,t+\Delta t]}g_e(\boldsymbol x-\boldsymbol R_e)$, with $\boldsymbol R_e$ the midpoint joining the two particles in pair-kick event $e$. We approximate the kicks by independent Poisson pair events and average over their lengths and directions. Following a similar approach used for lattice model in Ref.~\cite{Hexner2017}, we have
\begin{equation}
 \begin{aligned}
 \langle\Delta\tilde\rho(\boldsymbol q,t)\rangle
 &=-\tilde F(q)\tilde\rho(\boldsymbol q,t)\Delta t,\\
 \langle\Delta\tilde\rho(\boldsymbol q,t)\Delta\tilde\rho(\boldsymbol q',t)\rangle
 &=(2\pi)^2\tilde C(q)\delta(\boldsymbol q+\boldsymbol q')\Delta t,
 \end{aligned}
 \label{eq:cumulants}
\end{equation}
where the first relation determines diffusion, with the second specifying the noise statistics, and both relations are evaluated to the leading order $\Delta t$ in our calculations. Taking one simulation step as the unit of time, we use a local pair-event rate density $\rho_a\simeq\nu f_a\rho$, where $f_a$ is the active particle fraction and $\nu=N_{\mathrm{pair}}/N_{\mathrm{active}}$ accounts for multiple pair kicks per particle, with $N_{\mathrm{pair}}$ and $N_{\mathrm{active}}$ the numbers of overlapping pairs and active particles, respectively. We keep $\nu$ and $f_a$ at their mean values. The detailed calculation is given in the \textit{SI Appendix}.


Specifically, the first relation in Eq.~\ref{eq:cumulants} links the diffusion term $\tilde F(q)$ to the mean squared kick length, $\langle\ell^2\rangle_\ell\equiv\int_0^\infty d\ell\,\ell^2P(\ell)$. For $b>2$, this quantity remains finite, with $\langle\ell^2\rangle_\ell=b\sigma^2/[3(b-2)]$ obtained from Eq.~\ref{eq:pareto}. The leading density dynamics therefore remains ordinary diffusion, with $\tilde F(q)\simeq D_bq^2$ and $D_b=\nu f_a\langle\ell^2\rangle_\ell/2$ (see \textit{SI Appendix}). Physically, increasing $b$ makes long kicks less frequent and thus reduces $D_b$ at fixed $\nu f_a$, consistent with the qualitative trend of the particle diffusivity in Fig.~\ref{fig:kicks}(D).
The second relation in Eq.~\ref{eq:cumulants}, however, involves the squared density change caused by pair kick events, making the noise spectrum $\tilde C(q)$ more sensitive to long kick displacements. Consequently, its long-wavelength form depends on whether the fourth moment of the kick-length distribution $\langle\ell^4\rangle_\ell\equiv\int_0^\infty d\ell\,\ell^4P(\ell)$ obtained from Eq.~\ref{eq:pareto} is finite. For $2<b<4$, $\langle\ell^4\rangle_\ell$ diverges, and long kicks change the noise spectrum to $\tilde C(q)\simeq A_b^2q^b$ with noise strength parameter $A_b$ (see \textit{SI Appendix}). For $b>4$, $\langle\ell^4\rangle_\ell$ is finite, and the usual form $\tilde C(q)\simeq A_b^2q^4$ is recovered. At the marginal value $b=4$, the fourth moment diverges only logarithmically, so the noise spectrum retains the leading power $q^4$ with a logarithmic correction, $\tilde C(q)\propto q^4\ln[1/(q\sigma)]$.
Therefore, the density fluctuation field equation at large length scales for $2<b<4$ and $b>4$ is
\begin{equation}
 \begin{aligned}
 \partial_t\delta\rho&=D_b\nabla^2\delta\rho+\xi,\\
 \xi&=\nabla^2\zeta,\,\,\text{with}\,\,
 \zeta=A_b(-\nabla^2)^{\min(1,b/4)-1}\eta,
 \end{aligned}
 \label{eq:heavyfield}
\end{equation}
where $\eta$ is Gaussian white noise, $A_b^2=\nu f_a\bar\rho(2^{b+1}-8)C_b$ for $2<b<4$, with the coefficient $C_b$ calculated in \textit{SI Appendix}, and $A_b^2={3\nu f_a\bar\rho\,b\sigma^4}/{[40(b-4)]}$ for $b>4$. Therefore, for $b\ne4$, the driving noise takes the fractional-Laplacian form $\xi=-A_b(-\nabla^2)^{\min(1,b/4)}\eta$ with spectrum proportional to $q^{\min (b,4)}$, from which we obtain 
\begin{equation}
 S(q \to 0)\simeq\frac{A_b^2}{2\bar\rho D_b}q^{\min(b-2,2)}.
 \label{eq:heavySk}
\end{equation}
The case $b=4$ follows the quadratic scaling with a logarithmic correction, $S(q \to 0)\sim q^2\ln[1/(q\sigma)]$ (see \textit{SI Appendix}). 

These explain why the hyperuniform exponent decreases continuously as $b$ is reduced below $4$, while the motion remains diffusive. Intuitively, decreasing $b$ makes long jumps more frequent. Small reciprocal kicks mainly rearrange particles locally, so their effects largely cancel for the large-scale density fluctuations. On the other hand, long kicks can transfer particles between distant regions, generating stronger large-scale density fluctuations even though the center of mass remains conserved. This weakens the suppression of density fluctuations by reducing the hyperuniform exponent. Moreover, such long jumps also increase the diffusivity at a fixed kick rate, but remain sufficiently rare for particle motion to stay diffusive when $b>2$ in our case.

\section*{Discussion}
We have shown how spatially nonlocal noise can be used to design the hyperuniform scaling in center-of-mass-conserving BRO systems. Nonreciprocal coupling across dimensions transfers the density fluctuations of a three-dimensional Brownian gas to a two-dimensional particle layer, producing $S(q \to 0)\sim q^1$. Here the bulk geometry fixes the exponent to one, demonstrating that center-of-mass conservation does not restrict BRO to $q^2$ scaling. The effective planar description suggests coupling to gases with other dynamics, such as L\'evy flights, to obtain continuously tunable exponents, $S(q \to 0)\sim q^{2-\mu}$ with L\'evy parameter $0<\mu<2$. Heavy-tailed reciprocal kicks provide a second route, directly modifying the noise spectrum and producing $S(q \to 0)\sim q^{b-2}$ for $2< b < 4$. Within a temporally white description, both approaches realize spatially nonlocal driving noise of the fractional-Laplacian form $\xi\propto(-\nabla^2)^{(\alpha+2)/4}\eta$ , where $\eta$ is a unit Gaussian white noise, leading to $S(q)\sim q^\alpha$. The two protocols thus provide concrete ways to generate and control this nonlocal noise through particle dynamics, allowing the hyperuniform exponent to be controlled while the organized particles retain ordinary diffusion.



Our results further suggest possible ways to design hyperuniform scaling in nonequilibrium systems, such as bio-inspired sensing protocols, motivated by programmable quorum-sensing motility~\cite{Bauerle2018}, which could modify the spatial and temporal correlations generated by the density-sensing coupling. Besides, higher-order conservation laws could extend the accessible exponents beyond two, as demonstrated in lattice models~\cite{Maire2025}, although how to implement these rules in particle systems physically remains an open question. Moreover, temporal correlations provide another route~\cite{Ikeda2023,Ikeda2024,Anand2026}, since a power-law frequency spectrum of the driving noise can also tune hyperuniform scaling. It would therefore be interesting to design reciprocal-kick dynamics with controlled temporal correlations, allowing tunable hyperuniformity.

\matmethods{The two models use reciprocal kicks of overlapping particle pairs in a homogeneous active state. Every overlapping pair is updated once per BRO cycle, and a particle can participate in more than one pair event. Length is measured in units of the particle diameter $\sigma=1$, and time in units of the BRO update interval $\tau=1$. In the coupled model, the Brownian gas uses the same time step, $dt=\tau$, and the sensed density is evaluated around the midpoint of each overlapping pair. In the heavy-tailed model, kick lengths are independently drawn from Eq.~\ref{eq:pareto}. The structure factor and the particle MSD characterize density fluctuations and particle motion, respectively. The field theories use a linear approximation about a homogeneous active state; their Fourier conventions, noise correlations, and derivations are given in \textit{SI Appendix}.

}
\showmatmethods{} 

\dataavail{
Data will be made available upon reasonable request.}

\acknow{}
\showacknow{This work was financially supported by the Academic Research Fund from the Singapore Ministry of Education (RG151/23, and RG88/25) and the National Research Foundation, Singapore, under its 29th Competitive Research Program (CRP) Call (NRF-CRP29-2022-0002).} 

\bibsplit[23]

\bibliography{references}

\begin{thebibliography}{10}

\bibitem{Torquato2003}
S Torquato, FH Stillinger, Local density fluctuations, hyperuniformity, and
  order metrics.
\newblock {\em\protect\JournalTitle{Phys. Rev. E}} \textbf{68}, 041113 (2003).

\bibitem{Torquato2018}
S Torquato, Hyperuniform states of matter.
\newblock {\em\protect\JournalTitle{Phys. Rep.}} \textbf{745}, 1--95 (2018).

\bibitem{Jiao2014}
Y Jiao, et~al., Avian photoreceptor patterns represent a disordered
  hyperuniform solution to a multiscale packing problem.
\newblock {\em\protect\JournalTitle{Phys. Rev. E}} \textbf{89}, 022721 (2014).

\bibitem{Donev2005}
A Donev, FH Stillinger, S Torquato, Unexpected density fluctuations in jammed
  disordered sphere packings.
\newblock {\em\protect\JournalTitle{Phys. Rev. Lett.}} \textbf{95}, 090604
  (2005).

\bibitem{Hexner2015}
D Hexner, D Levine, Hyperuniformity of critical absorbing states.
\newblock {\em\protect\JournalTitle{Phys. Rev. Lett.}} \textbf{114}, 110602
  (2015).

\bibitem{Lei2019Chiral}
QL Lei, M Pica~Ciamarra, R Ni, Nonequilibrium strongly hyperuniform fluids of
  circle active particles with large local density fluctuations.
\newblock {\em\protect\JournalTitle{Sci. Adv.}} \textbf{5}, eaau7423 (2019).

\bibitem{Wilken2020}
S Wilken, RE Guerra, DJ Pine, PM Chaikin, Hyperuniform structures formed by
  shearing colloidal suspensions.
\newblock {\em\protect\JournalTitle{Phys. Rev. Lett.}} \textbf{125}, 148001
  (2020).

\bibitem{lei2023random}
Y Lei, N Zheng, R Ni, Random organization and non-equilibrium hyperuniform
  fluids on a sphere.
\newblock {\em\protect\JournalTitle{The Journal of Chemical Physics}}
  \textbf{159} (2023).

\bibitem{lei2023howHUfreeze}
Y Lei, R Ni, How does a hyperuniform fluid freeze?
\newblock {\em\protect\JournalTitle{Proc. Natl. Acad. Sci. U.S.A.}}
  \textbf{120}, e2312866120 (2023).

\bibitem{kuroda2023microscopic}
Y Kuroda, K Miyazaki, Microscopic theory for hyperuniformity in two-dimensional
  chiral active fluid.
\newblock {\em\protect\JournalTitle{J. Stat. Mech.: Theory Exp.}}
  \textbf{2023}, 103203 (2023).

\bibitem{Liu2024}
Y Liu, D Chen, J Tian, W Xu, Y Jiao, Universal hyperuniform organization in
  looped leaf vein networks.
\newblock {\em\protect\JournalTitle{Phys. Rev. Lett.}} \textbf{133}, 028401
  (2024).

\bibitem{zheng2024universal}
Y Zheng, MA Klatt, H L{\"o}wen, Universal hyperuniformity in active field
  theories.
\newblock {\em\protect\JournalTitle{Phys. Rev. Research}} \textbf{6}, L032056
  (2024).

\bibitem{li2025fluidization}
ZQ Li, QL Lei, YQ Ma, Fluidization and anomalous density fluctuations in {2D}
  {Voronoi} cell tissues with pulsating activity.
\newblock {\em\protect\JournalTitle{Proc. Natl. Acad. Sci. U.S.A.}}
  \textbf{122}, e2421518122 (2025).

\bibitem{lei2026metastable}
Y Lei, R Ni, Metastable hyperuniformity at discontinuous absorbing transitions
  (2026).

\bibitem{torquato2026hyperuniformity}
S Torquato, J Kim, MA Klatt, R Car, PJ Steinhardt, Hyperuniformity of weighted
  particle systems.
\newblock {\em\protect\JournalTitle{Physical Review X}} \textbf{16}, 011042
  (2026).

\bibitem{Florescu2009}
M Florescu, S Torquato, PJ Steinhardt, Designer disordered materials with
  large, complete photonic band gaps.
\newblock {\em\protect\JournalTitle{Proc. Natl. Acad. Sci. U.S.A.}}
  \textbf{106}, 20658--20663 (2009).

\bibitem{Leseur2016}
O Leseur, R Pierrat, R Carminati, High-density hyperuniform materials can be
  transparent.
\newblock {\em\protect\JournalTitle{Optica}} \textbf{3}, 763--767 (2016).

\bibitem{uche2006collective}
OU Uche, S Torquato, FH Stillinger, Collective coordinate control of density
  distributions.
\newblock {\em\protect\JournalTitle{Phys. Rev. E}} \textbf{74}, 031104 (2006).

\bibitem{ChenTorquato2018}
D Chen, S Torquato, Designing disordered hyperuniform two-phase materials with
  novel physical properties.
\newblock {\em\protect\JournalTitle{Acta Mater.}} \textbf{142}, 152--161
  (2018).

\bibitem{shi2023computational}
W Shi, D Keeney, D Chen, Y Jiao, S Torquato, Computational design of
  anisotropic stealthy hyperuniform composites with engineered directional
  scattering properties.
\newblock {\em\protect\JournalTitle{Phys. Rev. E}} \textbf{108}, 045306 (2023).

\bibitem{shi2025three}
W Shi, Y Jiao, S Torquato, Three-dimensional construction of hyperuniform,
  nonhyperuniform, and antihyperuniform disordered heterogeneous materials and
  their transport properties via spectral density functions.
\newblock {\em\protect\JournalTitle{Phys. Rev. E}} \textbf{111}, 035310 (2025).

\bibitem{Zhang2020Realizable}
G Zhang, S Torquato, Realizable hyperuniform and nonhyperuniform particle
  configurations with targeted spectral functions via effective pair
  interactions.
\newblock {\em\protect\JournalTitle{Physical Review E}} \textbf{101}, 032124
  (2020).

\bibitem{Wang2024Designer}
H Wang, S Torquato, Designer pair statistics of disordered many-particle
  systems with novel properties.
\newblock {\em\protect\JournalTitle{The Journal of Chemical Physics}}
  \textbf{160}, 044911 (2024).

\bibitem{Lomba2017}
E Lomba, JJ Weis, S Torquato, Disordered hyperuniformity in two-component
  nonadditive hard-disk plasmas.
\newblock {\em\protect\JournalTitle{Phys. Rev. E}} \textbf{96}, 062126 (2017).

\bibitem{Lei2019Hydro}
QL Lei, R Ni, Hydrodynamics of random-organizing hyperuniform fluids.
\newblock {\em\protect\JournalTitle{Proc. Natl. Acad. Sci. U.S.A.}}
  \textbf{116}, 22983--22989 (2019).

\bibitem{lei2025non}
Y Lei, R Ni, Non-equilibrium dynamic hyperuniform states.
\newblock {\em\protect\JournalTitle{Journal of Physics: Condensed Matter}}
  \textbf{37}, 023004 (2025).

\bibitem{Hexner2017}
D Hexner, D Levine, Noise, diffusion, and hyperuniformity.
\newblock {\em\protect\JournalTitle{Phys. Rev. Lett.}} \textbf{118}, 020601
  (2017).

\bibitem{Maire2025}
R Maire, L Chaix, Hyperuniformity and conservation laws in non-equilibrium
  systems.
\newblock {\em\protect\JournalTitle{J. Chem. Phys.}} \textbf{163}, 214507
  (2025).

\bibitem{Ikeda2024}
H Ikeda, Harmonic chain far from equilibrium: Single-file diffusion, long-range
  order, and hyperuniformity.
\newblock {\em\protect\JournalTitle{SciPost Phys.}} \textbf{17}, 103 (2024).

\bibitem{Huang2021}
M Huang, W Hu, S Yang, QX Liu, HP Zhang, Circular swimming motility and
  disordered hyperuniform state in an algae system.
\newblock {\em\protect\JournalTitle{Proc. Natl. Acad. Sci. U.S.A.}}
  \textbf{118}, e2100493118 (2021).

\bibitem{Oppenheimer2022}
N Oppenheimer, DB Stein, MY Ben~Zion, MJ Shelley, Hyperuniformity and phase
  enrichment in vortex and rotor assemblies.
\newblock {\em\protect\JournalTitle{Nat. Commun.}} \textbf{13}, 804 (2022).

\bibitem{pine2005chaos}
DJ Pine, JP Gollub, JF Brady, AM Leshansky, Chaos and threshold for
  irreversibility in sheared suspensions.
\newblock {\em\protect\JournalTitle{Nature (London)}} \textbf{438}, 997--1000
  (2005).

\bibitem{Corte2008}
L Cort{\'e}, PM Chaikin, JP Gollub, DJ Pine, Random organization in
  periodically driven systems.
\newblock {\em\protect\JournalTitle{Nat. Phys.}} \textbf{4}, 420--424 (2008).

\bibitem{Dean1996}
DS Dean, Langevin equation for the density of a system of interacting
  {Langevin} processes.
\newblock {\em\protect\JournalTitle{J. Phys. A: Math. Gen.}} \textbf{29},
  L613--L617 (1996).

\bibitem{Peng2026}
Y Peng, Y Lei, R Ni, Spatiotemporally correlated activity regulation of
  random-organizing hyperuniform states.
\newblock {\em\protect\JournalTitle{arXiv preprint arXiv:2610.21769}} (2026).

\bibitem{kwasnicki2017ten}
M Kwa{\'s}nicki, Ten equivalent definitions of the fractional laplace operator.
\newblock {\em\protect\JournalTitle{Fractional Calculus and Applied Analysis}}
  \textbf{20}, 7--51 (2017).

\bibitem{lischke2020fractional}
A Lischke, et~al., What is the fractional laplacian? a comparative review with
  new results.
\newblock {\em\protect\JournalTitle{Journal of Computational Physics}}
  \textbf{404}, 109009 (2020).

\bibitem{Chechkin2012}
AV Chechkin, IM Zaid, MA Lomholt, IM Sokolov, R Metzler, Bulk-mediated
  diffusion on a planar surface: Full solution.
\newblock {\em\protect\JournalTitle{Phys. Rev. E}} \textbf{86}, 041101 (2012).

\bibitem{stinga2017regularity}
PR Stinga, JL Torrea, Regularity theory and extension problem for fractional
  nonlocal parabolic equations and the master equation.
\newblock {\em\protect\JournalTitle{SIAM Journal on Mathematical Analysis}}
  \textbf{49}, 3893--3924 (2017).

\bibitem{Bychuk1995}
OV Bychuk, B O'Shaughnessy, Anomalous diffusion at liquid surfaces.
\newblock {\em\protect\JournalTitle{Phys. Rev. Lett.}} \textbf{74}, 1795--1798
  (1995).

\bibitem{Metzler2007}
R Metzler, AV Chechkin, VY Gonchar, J Klafter, Some fundamental aspects of
  {L{\'e}vy} flights.
\newblock {\em\protect\JournalTitle{Chaos, Solitons \& Fractals}} \textbf{34},
  129--142 (2007).

\bibitem{Fogedby1994}
HC Fogedby, Langevin equations for continuous time {L{\'e}vy} flights.
\newblock {\em\protect\JournalTitle{Phys. Rev. E}} \textbf{50}, 1657--1660
  (1994).

\bibitem{Jespersen1999}
S Jespersen, R Metzler, HC Fogedby, {L\'evy} flights in external force fields:
  Langevin and fractional fokker-planck equations and their solutions.
\newblock {\em\protect\JournalTitle{Physical Review E}} \textbf{59}, 2736--2745
  (1999).

\bibitem{Bauerle2018}
T B{\"a}uerle, A Fischer, T Speck, C Bechinger, Self-organization of active
  particles by quorum sensing rules.
\newblock {\em\protect\JournalTitle{Nat. Commun.}} \textbf{9}, 3232 (2018).

\bibitem{Ikeda2023}
H Ikeda, Correlated noise and critical dimensions.
\newblock {\em\protect\JournalTitle{Phys. Rev. E}} \textbf{108}, 064119 (2023).

\bibitem{Anand2026}
S Anand, G Zhang, S Martiniani, Spatial self-organization driven by temporal
  noise (2026).

\end{thebibliography}
\end{document}